\documentclass[3p]{elsarticle}

\usepackage{xcolor}
\usepackage{hyperref}       
\usepackage{url}            
\usepackage{booktabs}       
\usepackage{amsfonts}       
\usepackage{nicefrac}       
\usepackage{microtype}      
\usepackage{lipsum}
\usepackage{physics}
\usepackage{mathtools}
\usepackage{amsmath}
\usepackage{fancyhdr}       
\usepackage{graphicx}       
\usepackage{multirow}
\usepackage{caption}
\usepackage{subcaption}

\graphicspath{{media/}}     
\usepackage{color}

\biboptions{sort&compress}

\makeatletter
\def\ps@pprintTitle{%
  \let\@oddhead\@empty
  \let\@evenhead\@empty
  \def\@oddfoot{\reset@font\hfil\thepage\hfil}
  \let\@evenfoot\@oddfoot
}
\makeatother

\begin{document}

\begin{frontmatter}

\title{Quantum support vector data description for anomaly detection}

\author[1]{Hyeondo Oh}
\ead{leo9123@yonsei.ac.kr}
\address[1]{
Department of Statistics and Data Science, Yonsei University, Seoul, Republic of Korea
}

\author[1,2]{Daniel K. Park\corref{corr1}}
\ead{dkd.park@yonsei.ac.kr}
\address[2]{
Department of Applied Statistics, Yonsei University, Seoul, Republic of Korea
}
\cortext[corr1]{Corresponding author}

\begin{abstract}
Anomaly detection is a critical problem in data analysis and pattern recognition, finding applications in various domains. We introduce quantum support vector data description (QSVDD), an unsupervised learning algorithm designed for anomaly detection. QSVDD utilizes a shallow-depth quantum circuit to learn a minimum-volume hypersphere that tightly encloses normal data, tailored for the constraints of noisy intermediate-scale quantum (NISQ) computing. Simulation results on the MNIST and Fashion MNIST image datasets demonstrate that QSVDD outperforms both quantum autoencoder and deep learning-based approaches under similar training conditions. Notably, QSVDD offers the advantage of training an extremely small number of model parameters, which grows logarithmically with the number of input qubits. This enables efficient learning with a simple training landscape, presenting a compact quantum machine learning model with strong performance for anomaly detection.
\end{abstract}


\end{frontmatter}

\section{Introduction}
\label{sec:intro}

Quantum Machine Learning (QML) aims to overcome the limitations of classical counterparts in addressing various data analysis tasks by harnessing quantum information theory~\cite{10.1038/s43588-022-00311-3, 10.1080/00107514.2014.964942, 10.1038/nature23474}. Notably, QML algorithms achieved significant progress in the field of binary classification, a fundamental problem in pattern recognition. These algorithms demonstrate the potential to surpass classical methods in terms of runtime efficiency, trainability, model capacity, and prediction accuracy~\cite{10.1103/physrevlett.113.130503,10.1088/2058-9565/ab5944,cerezo2020variational,abbas_power_2021}.
Anomaly detection (AD) is another important branch of pattern recognition, spanning a wide range of applications, including finance~\cite{phua2010comprehensive,li2012identifying,jeragh2018combining}, bioinformatics~\cite{feher2014cell,min2017deep}, manufacturing~\cite{marti2015anomaly}, computer vision~\cite{6618951,bao2019computer}, and high energy physics~\cite{Fraser2022}. However, the task of constructing a machine learning model for AD is more intricate than binary classification due to the rarity of anomalies within sample data. Consequently, the training of AD models encounters limited label information, necessitating the adoption of unsupervised learning techniques~\cite{chandola2009anomaly, chalapathy2019deep}. A notable example of such techniques is one-class classification (OCC)~\cite{MOYA1996463, OCC_survey, OCC_survey2, OC-SVM, SVDD, DSVDD}, wherein a classifier is trained exclusively using a set of data samples that share the same (normal) class. After training, the one-class classifier determines whether the test data belongs to the same class as the training data. If not, the test data is classified as anomalous. 

OCC tasks have been addressed by statistical machine learning approaches, such as one-class support vector machine (OC-SVM)~\cite{OC-SVM}, support vector data description (SVDD)~\cite{SVDD}, and by deep learning-based algorithms~\cite{OCC_survey_deep, DSVDD}. SVDD is a kernel-based approach with an optimization procedure that can be interpreted as identifying the smallest hypersphere that exclusively encloses normal data. The test data that lies outside of the hypersphere is classified as anomalous. Deep SVDD (DSVDD) further enhances the SVDD technique by identifying the feature space where the separation between normal and abnormal data is maximized through deep learning~\cite{DSVDD}. Recent advances in QML has motivated the development of quantum AD techniques to enhance existing classical methods~\cite{QML_for_AD, QAD_audio, QAD_phase_diagram, VQOCC}. However, several challenges remain to be addressed. For instance, the quantum algorithms proposed in~\cite{QML_for_AD} promise exponential speedup in theory but rely on expensive subroutines such as the quantum linear solver~\cite{HHL} and matrix exponentiation~\cite{qPCA}, which are not feasible for Noisy Intermediate-Scale Quantum (NISQ) computing~\cite{NISQ}. Moreover, these algorithms are tailored for quantum data, which limits their applicability when the input data is classical. Quantum algorithms more suited for NISQ devices focus on applying quantum autoencoders (QAE)~\cite{romero2017quantum,bravo2021quantum,VQOCC,AD_qae,AD_qae_hybrid}, which is optimized for data compression, to AD, rather than directly addressing the AD objective.

To address these limitations, we propose quantum support vector data description (QSVDD), a shallow-depth variational QML algorithm that is purposefully built for AD and tailored for NISQ computing. Drawing inspiration from DVSDD and the variational quantum algorithm (VQA)~\cite{cerezo2021variational, blance2021quantum, chen2020variational, romero2021variational}, QSVDD utilizes the variational quantum circuit (VQC) to learn useful feature representations of the data together with the objective inherited from SVDD. Within our framework, the VQC employs the quantum convolutional neural network (QCNN) architecture~\cite{cong_QCNN}. This implementation enables logarithmic growth in both circuit depth and the number of parameters relative to the number of input qubits. The former property is particularly beneficial for NISQ devices, in which the size of quantum circuits that can be executed reliably is limited. Meanwhile, the latter attribute positions QSVDD as an extremely compact machine learning model. In addition, the absence of barren plateaus which guarantees trainability~\cite{pesah2020absence}, good generalization capabilities~\cite{PRXQuantum.2.040321}, and the excellent binary classification performance demonstrated across various tasks~\cite{cong_QCNN, hur_QCNN, kim2023classical} make QCNN the favorable choice.

Numerical experiments were conducted on the MNIST and Fashion MNIST image datasets using Pennylane~\cite{bergholm2020pennylane}. Each class within the dataset was treated as normal, resulting in ten independent AD tasks. The performance of each method was assessed through the area under the receiver operating characteristic (ROC) curve (AUC). For all ten AD tasks, QSVDD outperformed both QAE-based method and the classical DSVDD under similar training conditions. QSVDD achieves these results while learning fewer parameters than QAE and DSVDD, as well as employing a shallower circuit than QAE. This establishes QSVDD as an efficient and effective algorithm for AD.

The remainder of the paper is organized as follows. Section~\ref{sec:prelim} briefly reviews OCC, SVDD, and DSVDD to establish the background for QSVDD. Section~\ref{sec:qsvdd} presents the QSVDD method, explaining its main components in detail. Section~\ref{sec:numexp} describes the numerical experiments conducted to evaluate the performance of QSVDD and compare it to existing methods, such as QAE and DSVDD. Conclusions and suggestions for future research directions are provided in Section~\ref{sec:conc}.

\section{Related Work} \label{sec:prelim}

Before introducing QSVDD, we briefly describe OCC to set up the problem. Following this, we review SVDD and DSVDD, two widely recognized methods for OCC that serve as the basis upon which we develop QSVDD.

\subsection{One-class classification}
\label{sec:occ}
The goal of classification in general is to predict the class label of a given test dataset denoted as $\tilde{\boldsymbol{x}}\in\mathbb{R}^N$, based on a labeled training dataset. Formally, a labeled training dataset can be represented as:
\begin{equation} \label{eq: prelim_dataset}
\mathcal{D} = \{ (\boldsymbol{x}_1,y_1), \dots, (\boldsymbol{x}_m,y_m) \} \subset \mathbb{R}^N \times \mathbb{Z}_L,
\end{equation}
where $L$ is the total number of classes within the dataset. The problem is commonly referred to as binary classification when $L=2$, and as multi-class classification when $L>2$. 

One-class classification (OCC) involves datasets with $L=1$, indicating that all data samples share the same label value. As a result, the dataset defined in Eq.~(\ref{eq: prelim_dataset}) is reduced to $\mathcal{D}_{1} = \{ \boldsymbol{x}_1, \dots, \boldsymbol{x}_m \} \subset \mathbb{R}^N$. The goal of OCC is to identify data points that belong to the same class as the training data with a high success probability~\cite{OCC_survey, OCC_survey2}. In the context of AD, the data samples contained in $\mathcal{D}_{1}$ are regarded as representing normal instances. Thus, the one-class classifier decides whether the test data belongs to the normal set or is anomalous. The OCC algorithm accomplishes this by constructing a decision function, denoted by $f(\boldsymbol{x};\mathcal{D}_1)$, which quantifies the degree of dissimilarity between the test data $\tilde{\boldsymbol{x}}$ and the training dataset. The decision rule is established by utilizing a predefined threshold $b$ as follows: If $f(\tilde{\boldsymbol{x}};\mathcal{D}_1)>b$, then $\tilde{\boldsymbol{x}}$ is classified as an anomaly. Conversely, if $f(\tilde{\boldsymbol{x}};\mathcal{D}_1) < b$, the test data is accepted as normal. If $f(\tilde{\boldsymbol{x}};\mathcal{D}_1) = b$, then the decision can be made at random. The design of an appropriate decision function holds fundamental importance in machine learning, as it significantly influences overall performance~\cite{jain2000statistical}.

\subsection{Support vector data description}
\label{sec:svdd}
Support vector data description (SVDD)~\cite{SVDD} can be thought of as a derivative of the support vector machine (SVM)~\cite{SVM}, tailored for OCC. While SVM seeks a maximum margin hyperplane that separates two classes, the objective of SVDD is to make a description of a training dataset by finding a hypersphere that encompasses the dataset. To minimize the chance of including data that differs from the target training samples, the volume of the hypersphere is minimized. The training data points located on the boundary of the hypersphere are known as support vectors. Consequently, these support vectors succinctly describe the entire dataset. In the mathematical formulation of SVDD, the data points only appear in the form of inner products. This property permits the application of the kernel trick~\cite{hofmann2006support} to effectively handle nonlinear data patterns. SVDD is particularly beneficial when dealing with limited data resources, which is common in most AD scenarios~\cite{tax2002one, SVDD}, as it provides a data description without the need to estimate data density.

The decision function of SVDD is expressed as
\begin{equation} \label{eq: decision_SVDD}
f(\boldsymbol{x}) = ||\Phi(\boldsymbol{x}) - \mathbf{c}||^2 - r,
\end{equation}
with the threshold $b=0$. In this formulation, $\Phi:\mathbb{R}^{d}\rightarrow \mathbb{R}^{d'}$ represents the feature mapping, while $\boldsymbol{c}$ and $r$ denote the center and the radius of the hypersphere, respectively. The optimization process for SVDD seeks to determine optimal values for $\boldsymbol{c}$ and $r$ in a way that minimizes the volume of the hypersphere enclosing the normal data. Consequently, the decision function identifies data points lying outside this boundary as anomalies.

Note that when all data is normalized to unit norm vectors, SVDD is equivalent to the one-class SVM~\cite{tax2002one} which aims to find a maximum margin hyperplane that best separates the training dataset from the origin~\cite{10.1162/089976601750264965}.

\subsection{Deep support vector data description}
\label{sec:dsvdd}

Kernel-based approaches, such as SVDD, face several drawbacks. For instance, they are sensitive to the choice of the kernel function, requiring explicit feature engineering tailored to the specific dataset~\cite{smola1998learning, pal2010feature, DSVDD}. Moreover, the construction and manipulation of the kernel matrix results in  poor computational scaling, with a complexity of $O(m^2)$ for $m$ training samples~\cite{vempati2010generalized}. Furthermore, prediction using kernel methods mandates the storage of support vectors, a requirement that can consume substantial memory resources. Deep support vector data description (DSVDD) overcomes these limitations by integrating deep learning. The primary modification involves implementing the feature mapping through a neural network equipped with a set of trainable parameters denoted as $\boldsymbol{\theta}$. The neural network is trained with the one-class classification objective, ensuring that the learning process corresponds to the task of finding the optimal feature space for the data and the minimum-volume hypersphere.

The decision function of DSVDD can be expressed as
\begin{equation} \label{eq: decision_DSVDD}
f(\boldsymbol{x}) = ||\Phi(\boldsymbol{x},\boldsymbol{\theta}) - \mathbf{c}||^2 - r,
\end{equation}
where $r$ is the radius of the hypersphere and the threshold is set to $b=0$. Here, $\Phi:\mathbb{R}^{d}\rightarrow \mathbb{R}^{d'}$ represents the neural network responsible for mapping the original data into the feature space. The neural network comprises $h\in\mathbb{N}$ hidden layers and a set of weights $\boldsymbol{\theta}=\{ \boldsymbol{\theta}^1, \dots, \boldsymbol{\theta}^h \}$. Typically, the number of output nodes in the neural network is fewer than the input nodes, i.e. $d' < d$, aiming to eliminate redundant information and reduce computational complexity. Similar to SVDD, $\boldsymbol{c}$ and $r$ are the center and the radius of the hypersphere, respectively. Given a training set of normal data $\lbrace \boldsymbol{x}_i \rbrace_{i=1,\ldots,m}$, the loss function subject to minimization is defined as
\begin{equation} \label{eq: svdd}
L_c(\boldsymbol{\theta}) = \frac{1}{m}\sum_{i=1}^{m}|| \Phi(\boldsymbol{x}_i, \boldsymbol{\theta}) - \boldsymbol{c} ||^2 + \lambda R(\boldsymbol{\theta}).
\end{equation}
In this equation, the second term represents the regularization term with a hyperparameter $\lambda>0$, added to prevent overfitting. 

The center of the hypersphere $\boldsymbol{c}$ can be constructed by employing a deep convolutional autoencoder (DCAE). The network architecture of the encoder is identical to that of the DSVDD, while the decoder is designed symmetrically with the exception of using upsampling instead of max-pooling. After training the DCAE with the mean squared error loss, the hypersphere center is determined to be the mean of the vectors obtained through the encoder using a subset of training data samples (e.g. ten percent of the entire data). Empirical studies have demonstrated the effectiveness of this strategy~\cite{DSVDD}.

\section{Quantum support vector data description}
\label{sec:qsvdd}

\subsection{Overview} 
\label{sec:overview}

Quantum Support Vector Data Description (QSVDD) draws inspiration from DSVDD. However, QSVDD utilizes quantum computing techniques to optimize the feature map, $\Phi:\mathbb{R}^{d}\rightarrow \mathbb{R}^{d'}$. The potential advantage of this approach lies in the ability of quantum computers to efficiently manipulate data encoded within a quantum Hilbert space, which grows exponentially with the number of qubits~\cite{havlivcek2019supervised, schuld2019quantum, Huang_2021}. To be more specific, the feature mapping in QSVDD is given by $\Phi(\boldsymbol{x},\boldsymbol{\theta}) = g(U(\boldsymbol{\theta})|\psi(\boldsymbol{x})\rangle)$. This function begins with the process of quantum data encoding~\cite{havlivcek2019supervised,schuld2019quantum,PhysRevA.101.032308,PhysRevA.102.032420,araujo_divide-and-conquer_2021,araujo2021configurable}, denoted by $|\psi(\boldsymbol{x})\rangle$, which maps input data to a quantum state in the exponentially large Hilbert space. It also involves a variational quantum circuit (VQC) represented as $U(\boldsymbol{\theta})$, and $g:\mathbb{C}^{2^n}\rightarrow \mathbb{R}^{d'}$ acts as an embedding function that projects into a lower-dimensional space to ensure $d'<d$. While comprehensive discussions of these individual components are deferred to subsequent sections, it is important to highlight that QSVDD provides flexibility in selecting these components. This freedom allows for customization of the algorithm to address specific tasks effectively. The concept of transitioning from the high-dimensional quantum Hilbert space to a low-dimensional space resembles the projected quantum kernel, which has been demonstrated to achieve better prediction accuracy than classical machine learning models for quantum learning problems~\cite{Huang_2021}. 

Since the trainable parameters of a quantum circuit typically range from 0 to $2\pi$, the regularization term is unnecessary. Therefore, the loss function to be minimized in QSVDD can be expressed as
\begin{equation} \label{eq: qsvdd}
L_q(\boldsymbol{\theta}) = \frac{1}{m}\sum_{i=1}^{m}|| \Phi(\boldsymbol{x}_i, \boldsymbol{\theta}) - \mathbf{c} ||^2,
\end{equation}
where $\lbrace \boldsymbol{x}_i \rbrace_{i=1,\ldots,m}$ is the set of normal data samples and $\boldsymbol{c}$ represents the center of the hypersphere as before. Here, the VQC is trained with the one-class classification objective, akin to the training process of a neural network in DSVDD. Minimizing the loss function corresponds to learning the data representation in the quantum Hilbert space, such that the normal data points converge tightly towards the center of the hypersphere. The minimization of the QSVDD loss function in Equation~\ref{eq: qsvdd} can be carried out by using a classical optimization technique, such as the stochastic gradient descent. 

The center of the hypersphere $\mathbf{c}$ in QSVDD can be determined in a manner similar to that in DSVDD, with the exception of employing QAE instead of DCAE. Alternatively, one can opt for using a constant vector, such as $\boldsymbol{0}$. We compared the performance of QSVDD using these approaches on the MNIST and Fashion MNIST datasets and did not observe a significant difference. However, the former approach consumes more computational resources. Therefore, our empirical studies indicate that setting the center as $\boldsymbol{c}=\boldsymbol{0}$ is a practical and effective strategy.

When testing an unseen data $\tilde{\boldsymbol{x}}$, the following decision function is used:
\begin{equation} \label{eq: decision_QSVDD}
f(\tilde{\boldsymbol{x}}) = || \Phi(\tilde{\boldsymbol{x}}, \arg\min_{\boldsymbol{\theta}}L(\boldsymbol{\theta})) - \boldsymbol{c} ||^2 - r,
\end{equation}
where $r$ is the radius of the hypersphere and the threshold is set to $b=0$. If the value of the decision function for a test data point surpasses this threshold $b$, it is recognized as an anomaly; conversely, if it falls below the threshold, it is considered normal.

\begin{figure*}[t]
    \centering
    \includegraphics[width=\textwidth]{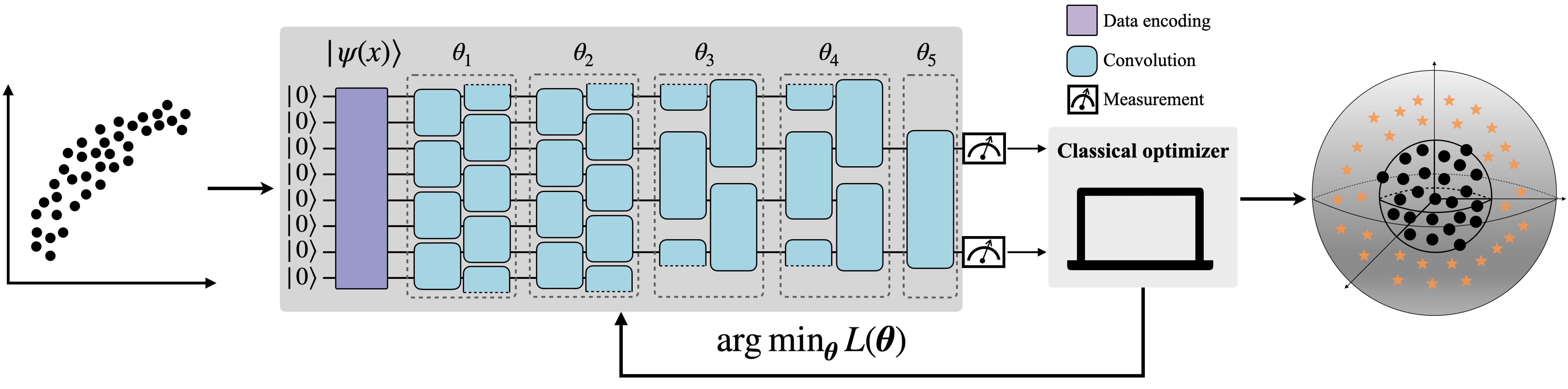}
    \caption{\label{fig1}QSVDD consists of four main components: data encoding, variational quantum circuit, measurement, and optimization. The classical input data is encoded into quantum data with $n$ qubits, where $n$ depends on the number of features. This encoding produces the state $|\psi(x)\rangle$. Subsequently, the quantum data undergoes a QCNN circuit. In the circuit diagram, the dashed line on the convolutional gate indicates its connection through the top and bottom wires to fulfill the translational invariance and the periodic boundary condition. During the measurement process, the expectation values of Pauli observables are computed, and these values represent the engineered features projected to a low-dimensional latent space. Moreover, these values are used in the optimization step. The trainable parameters within the variational quantum circuit are optimized using a classical optimizer to minimize the loss function in Eq.~(\ref{eq: qsvdd}). The optimization process seeks to construct the minimum-volume hypersphere that encompasses the normal data points within the latent space. The support vectors of QSVDD are represented by the points on the surface of the inner hypersphere. These boundary data points play a pivotal role in accurately describing the shape and characteristics of the hypersphere. By effectively capturing the compact hypersphere, anomalies (depicted as yellow stars) that exist outside its boundaries can be detected.}
\end{figure*}

The general concept of a QSVDD for application to classical data is illustrated in Fig.~\ref{fig1}. When the data is already provided as a quantum state from a quantum device, the QSVDD algorithm skips the data encoding part.

\subsection{Data encoding} \label{sec:data_enc}

QML requires that the dataset is given in the form of a quantum state. Thus, to perform a QML algorithm on classical data, classical data must be first mapped to a quantum state using a quantum feature map, which can be represented by the function $\Psi: \mathcal{X} \rightarrow \mathcal{H}$, where $\mathcal{X}$ is the original data space and $\mathcal{H}$ is a quantum Hilbert space~\cite{schuld2019quantum,lloyd2020quantum,hur_QCNN}. Encoding classical data $x$ as an $n$-qubit quantum state $|\psi(x) \rangle \in \mathbb{C}^{2^n}$ can be achieved by defining a unitary transformation $U_\psi (x)$ as a function of $x$ and applying it to a fiducial state, such as $|0\rangle^{\otimes n}$. In this case, the quantum feature map can be expressed $x\in \mathcal{X} \mapsto|\psi(x) \rangle = U_\psi (x) |0\rangle^{\otimes n}\in \mathcal{H} $.

QSVDD is compatible with any quantum feature map~\cite{PhysRevA.102.032420, havlivcek2019supervised, araujo_divide-and-conquer_2021,araujo2021configurable}. In this work, we focus on amplitude encoding as an example to deliver the main idea. Amplitude encoding represents an $N$-dimensional vector $x=(x_1, \dots, x_N)^{\top}$ as the probability amplitudes of an $n$-qubit quantum state as
\begin{equation} \label{eq: amp_enc}
| \psi(x) \rangle = \frac{1}{||x||}\sum_{i=1}^{N} x_i |i \rangle, 
\end{equation}
where $N=2^n$, and $|i\rangle$ corresponds to the $i$th computational basis state. 

The depth and the number of parameters of a VQC grow at most polynomially with the number of input qubits~\cite{cerezo2020variational}. This implies that amplitude encoding permits an exponential reduction in the count of trainable parameters. Some VQCs, such as the quantum convolutional neural network (QCNN)~\cite{cong_QCNN,hur_QCNN,kim2023classical}, can be constructed with both depth and the number of parameters growing logarithmically in relation to the number of input qubits. With amplitude encoding, this logarithmic scaling achieves a double-exponential reduction in the number of trainable parameters, thus constituting the most compact QML model.

When the task at hand involves detecting anomalies in quantum data generated by quantum systems, the data encoding step becomes unnecessary. Moreover, the quantum advantage of QSVDD is highly likely to be achieved in this scenario, as the process of extracting classical information from a quantum state, which is subsequently input to a machine learning model, is generally computationally demanding~\cite{10.1038/s43588-022-00311-3}.

\subsection{Variational quantum circuit}\label{sec:qsvdd_vqc}
The VQC is the sole trainable component within QSVDD, responsible for learning valuable feature representations of the data in the Hilbert space based on the given data encoding and projection function. Crucial aspects to be considered when designing the structure of the parameterized quantum circuit, also known as the ansatz, are the absence of barren plateaus~\cite{McClean2018} and the circuit depth. The former ensures the feasibility for training, while the latter addresses both computational complexity and resilience to noise. While QSVDD is inherently compatible with any VQCs and ansatzes, the QCNN structure stands out for several reasons. Notably, its hierarchical structure circumvents the issue of barren plateaus~\cite{pesah2020absence}, and the quantum circuit depth grows logarithmically in proportion to the number of input qubits. Moreover, an information-theoretic analysis underscores the strong generalization capabilities of QCNN~\cite{PRXQuantum.2.040321}. Furthermore, machine learning algorithms based on QCNN have demonstrated excellent performance in binary classification tasks involving both quantum and classical data~\cite{maccormack_branching_2020,hur_QCNN, kim2023classical}. Consequently, we focus on QCNN as our choice for VQC.

\begin{figure*}[t]
    \centering
    \includegraphics[width=0.6\textwidth]{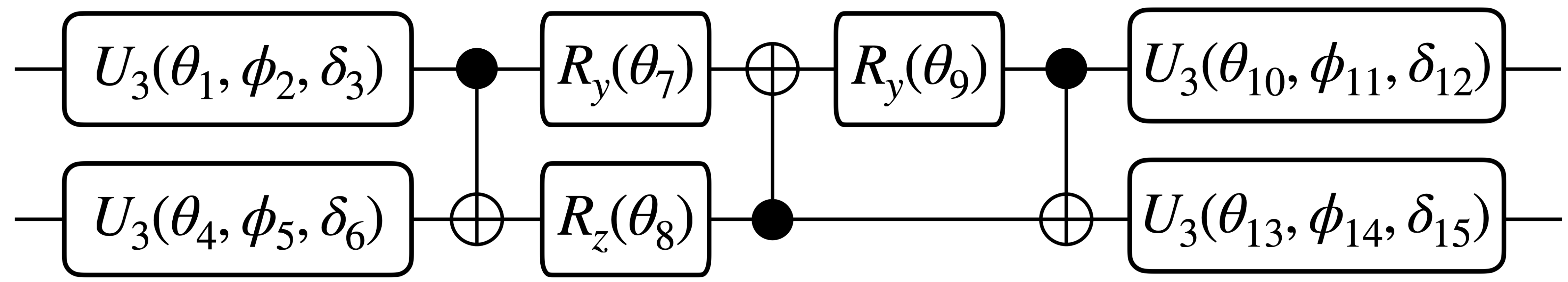}
    \caption{\label{fig2}The two-qubit circuit capable of implementing an arbitrary two-qubit unitary transformation in $SU(4)$. Here, $R_i(\theta)$ represents a single-qubit rotation around the $i$-axis of the Bloch sphere by an angle $\theta$, and $U_3(\theta, \phi, \delta) = R_z(\phi)R_x(-\pi/2)R_z(\theta)R_x(\pi/2)R_z(\delta)$. Furthermore, the circuit includes the controlled-NOT gate. The total number of parameters is fifteen. This circuit is utilized as the quantum convolution operation in this study.}
\end{figure*}

The QCNN comprises quantum convolution and pooling layers~\cite{cong_QCNN}, inspired by classical convolutional neural networks (CNNs). The quantum convolution layer exhibits translational invariance, meaning that the quantum circuit remains unchanged under the translation of qubits. Typically, it consists of parameterized two-qubit gates between the nearest neighboring qubits with a periodic boundary condition. Translational invariance implies that parameter sharing, a technique commonly employed to mitigate overfitting and reduce computational resources in classical CNN~\cite{liu2017survey}, is applied and all two-qubit gates within the same convolution layer share identical parameter values. While the two-qubit gate can be parameterized in various ways, an arbitrary $SU(4)$ operation can be generated using a circuit shown in Fig.~\ref{fig2} with fifteen real parameters. Quantum pooling in QCNN reduces the number of active qubits by a factor of two, facilitating the feature subsampling through the partial trace operation. Consequently, the depth of a QCNN circuit is in $O(\log(n))$ for $n$ input qubits. Due to parameter sharing and the logarithmic circuit depth, a QCNN is characterized by $O(\log(n))$ parameters. The quantum pooling operation may include parameterized two-qubit controlled-unitary gates prior to the dimensionality reduction, with control qubits being traced out after gate operations. Quantum convolution and pooling layers are iterated until the remaining system size is sufficiently small. For instance, the iteration can continue until only one qubit remains for binary classification~\cite{hur_QCNN}.

An example circuit of QCNN with eight input qubits is shown in Fig.~\ref{fig1}. The circuit begins with the data encoding step that prepares the input data as a quantum state. In this specific case, two convolution layers precede the pooling layer, and the sequence of convolution-convolution-pooling layers is iterated twice, culminating in a final state with two remaining qubits. These qubits subsequently undergo an additional layer of convolutional operations, succeeded by two-qubit measurements. The set of trainable parameters can be grouped into five, each corresponding to a quantum convolution layer, as $\boldsymbol{\theta} = \{\boldsymbol{\theta}_1, \boldsymbol{\theta}_2, \boldsymbol{\theta}_3, \boldsymbol{\theta}_4, \boldsymbol{\theta}_5 \}$. Since each parameter vector is associated with one convolution layer, $\mathrm{dim}(\boldsymbol{\theta}_i)\le 15 $ for all $i$. The rounded rectangles within the circuit diagram symbolize the two-qubit unitary gate, applied to nearest neighbor qubits. This operation is also extended to the top and bottom qubits to satisfy the periodic boundary condition. 

The circuit that enables the generation of an arbitrary two-qubit unitary transformation is shown in Fig.~\ref{fig2}. In the figure, $R_i(\theta)=\cos(\theta/2)I-i\sin(\theta/2)\sigma_i$ represents the single-qubit rotation gate around the $i$-axis of the Bloch sphere by an angle $\theta$, where $I$ is the 2-by-2 identity matrix and $\sigma_i$ with $i\in \lbrace x,y,z\rbrace$ is the Pauli matrix. Moreover, $U_3(\theta, \phi, \delta) = R_z(\phi)R_x(-\pi/2)R_z(\theta)R_x(\pi/2)R_z(\delta)$ is capable of generating an arbitrary single-qubit unitary transformation. The circuit also uses the controlled-NOT (CNOT) operation, which flips the state of the target qubit if the control qubit is in the state $|1\rangle$. In Fig.~\ref{fig1}, the quantum pooling layer only performs the partial trace operation without incorporating parameterized gates. This is particularly useful when the convolution layer already uses the parameterized gate in the form of Fig.~\ref{fig2}, which is capable of learning arbitrary two-qubit operations. In such instances, introducing parameterized gates to the pooling layers may not significantly alter the classification performance~\cite{hur_QCNN}. Therefore, considering computational efficiency, it can be preferable to leave the quantum pooling layer unparameterized when the convolution layer is parameterized using the $SU(4)$ circuit.

If the nature of the target problem necessitates a significant increase in model complexity by utilizing a larger number of parameters, it is possible to stack numerous convolution layers before each pooling operation. Deactivating parameter sharing is an alternative strategy, permitting each two-qubit gate to independently take distinct parameter values~\cite{grant_hierarchical_2018}.

\subsection{Measurement}
\label{sec:qsvdd_meas}
The data encoded in the quantum Hilbert space and manipulated by the quantum circuit gets embedded into the final low-dimensional feature (i.e. latent) space through a function $g:\mathbb{C}^{2^n}\rightarrow \mathbb{R}^{d'}$. As discussed in Section~\ref{sec:dsvdd}, it is imperative to eliminate redundant information and reduce the computational complexity for both computing and minimizing the loss function. Consequently, the condition $d' < d$ is imposed. Adopting a QCNN as the VQC in QSVDD as described in the previous section naturally achieves dimensionality reduction. Since QCNN involves entangling gates and partial trace operations, an initial $n$-qubit pure state at the input transforms into a final $n_o$-qubit mixed state at the output. The former and latter states are fully characterized by $2^n-1$ and $4^{n_o}-1$ parameters, respectively, accounting for the normalization condition. To ensure that the output feature space is smaller than the input space, a minimum of two pooling layers is required. This leads to $n_o<n/2$, satisfying the condition $4^{n_o} < 2^n$. The QCNN circuit example illustrated in Fig.~\ref{fig2} applies two pooling layers, resulting in a two-qubit reduced density matrix at the output. The fifteen real values that completely characterize the two-qubit density matrix can be extracted by measuring the expectation values of two-qubit Pauli observables. Specifically, the $i$th value is evaluated as $v_i=\langle P_i\rangle$, where $P_i \in \mathcal{P}_2 = \lbrace I, \sigma_x, \sigma_y, \sigma_z\rbrace^{\otimes 2}$ and $\mathcal{P}_2$ denotes the set of two-qubit Pauli operators that forms a basis for the real vector space of two-qubit density matrices. Thus, the maximum dimension of the feature space in this particular example is fifteen. However, this does not mean that the size of the feature space always has to be $4^{n_o}-1$. It can be further reduced by measuring Pauli observables from a subset $\mathcal{S}\subset\mathcal{P}_2$, where $\vert \mathcal{S}\vert < \vert \mathcal{P}_2\vert$. Therefore, when $n_o=2$, the size of the feature space can range from one to fifteen. The dimension of the feature space can be adjusted to strike the optimal balance between accuracy and computational cost.

Note that the QCNN circuit shown in Fig.~\ref{fig1} uses a convolution layer before performing the expectation value measurement. The expectation value of a Pauli observable $P_i$ can be expressed as
\begin{equation}
    \label{eq:expval}
    \langle P_i\rangle = \mathrm{Tr}\left(P_iU_c(\boldsymbol{\theta}_5)\rho_2 U_c^{\dagger}(\boldsymbol{\theta}_5)\right) = \mathrm{Tr}\left(\tilde{P}_i(\boldsymbol{\theta}_5)\rho_2\right),
\end{equation}
where $U_c$ represents the two-qubit unitary gate in the convolution layer with the set of parameters $\boldsymbol{\theta}_5$ taking the two-qubit reduced matrix, denoted by $\rho_2$, as the input, and $\tilde{P}_i(\boldsymbol{\theta}_5)=U_c^{\dagger}(\boldsymbol{\theta}_5)P_iU_c(\boldsymbol{\theta}_5)$. The last expression is obtained by using the cyclic property of the trace operation, represented as $\mathrm{Tr}(ABC) = \mathrm{Tr}(CAB)$. Equation~(\ref{eq:expval}) shows that the measurement process and the last convolution layer can collectively be seen as a parameterized measurement. Therefore, optimizing the final convolution layer corresponds to the process of learning the optimal set of measurements to engineer features projected into the low-dimensional feature space.

\section{Numerical experiments} \label{sec:numexp}

The numerical experiments are conducted on the MNIST dataset and the Fashion MNIST dataset with eight input qubits using PennyLane~\cite{bergholm2020pennylane}. The images are resized from their original dimension of $28 \times 28 = 784$ to $16 \times 16 = 2^8$ using the bilinear interpolation method, a commonly used technique for image resizing~\cite{han2013comparison}, to facilitate amplitude encoding for loading the data to eight input qubits.

Both datasets contain ten distinct classes, facilitating the establishment of ten independent AD tests for each dataset. Each test considers one of the classes as the normal class. During each test, the QSVDD is trained using approximately 6000 normal data samples. Each of these tests was repeated five times, with parameters of the VQC being initialized randomly each time. The parameter initialization is carried out by sampling from a standard normal distribution. Subsequently, an Adam optimizer~\cite{kingma2014adam} is employed to update these parameters iteratively, aiming to minimize the loss function specified in Eq.~(\ref{eq: qsvdd}). Following the training phase, the area under the receiver operating characteristic (ROC) curve (AUC) is computed as the figure of merit. The AUC is evaluated using 1000 normal samples and 900 abnormal samples, consisting of 100 samples randomly selected from each of the remaining classes.

\begin{figure*}[t]
    \centering
    \includegraphics[width=.65\linewidth]{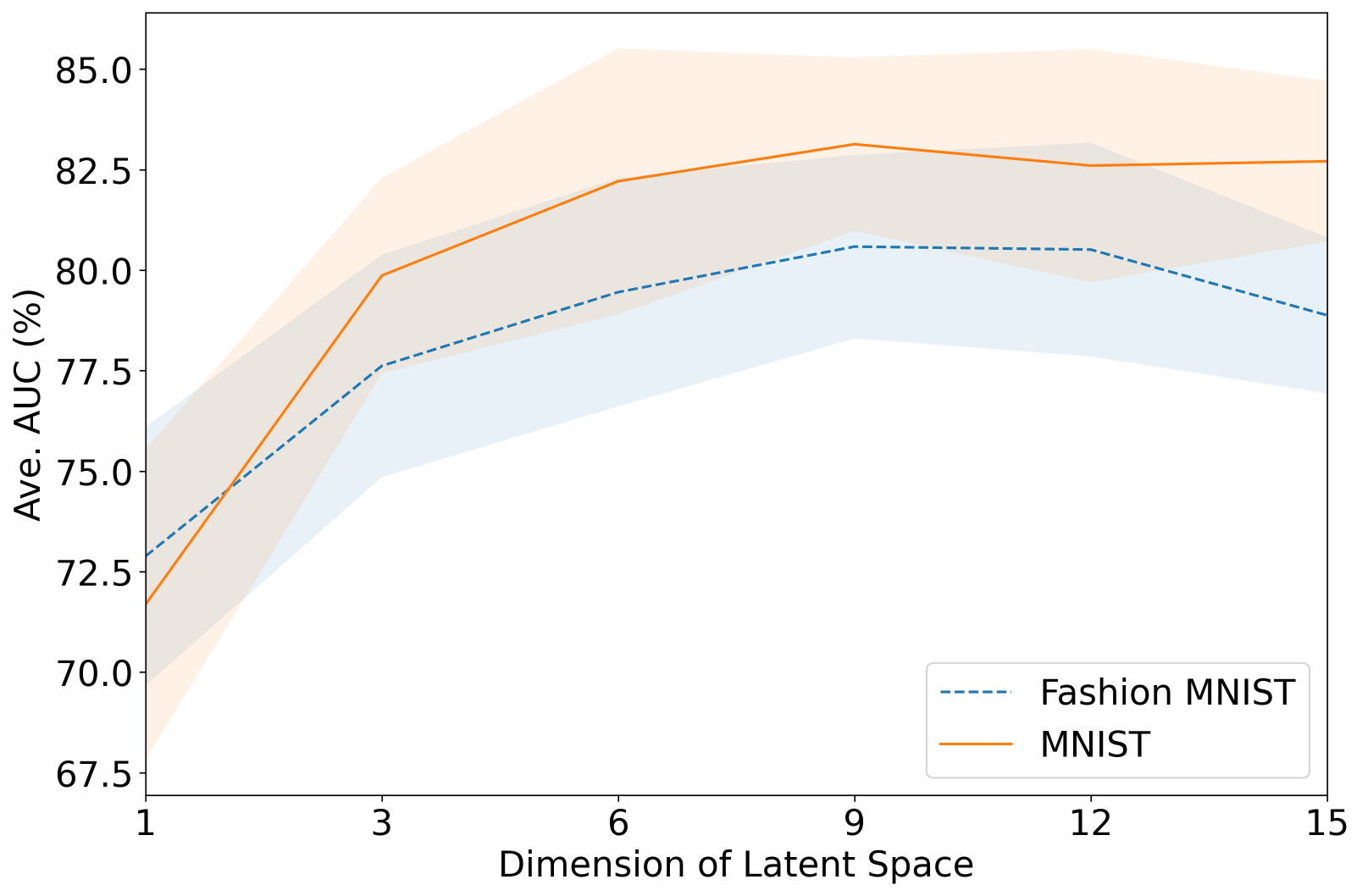}
    \caption{The average AUC scores for the MNIST dataset (solid line) and the Fashion MNIST dataset (dashed line) as a function of the latent space dimension, which corresponds to the number of two-qubit Pauli observables. For each dataset, the average is computed based on five repetitions of ten anomaly detection (AD) tests, each of which designates one of the classes as the normal class. The shaded areas correspond to standard deviations. The optimal latent space size for both datasets is determined to be nine.
    } %
    \label{AUC_mean}
\end{figure*}
Figure~\ref{AUC_mean} presents the AUC scores averaged over all ten AD tests with five random seeds as a function of the latent space dimension, determined by the number of two-qubit Pauli observables measured at the end of the VQC. The solid and dashed lines represent the results for the MNIST and Fashion-MNIST datasets, respectively, with shaded areas around each line indicating the corresponding standard deviation. The results demonstrate that QSVDD's performance steadily improves as the dimension of the latent space increases until reaching nine, beyond which there is no significant improvement. Due to the increasing computational time associated with a larger number of Pauli observables, we fix the latent space dimension at nine for all subsequent simulations.

\begin{figure}[t]
\begin{subfigure}[b]{.5\textwidth}
  \centering
  \includegraphics[height=5.5cm]{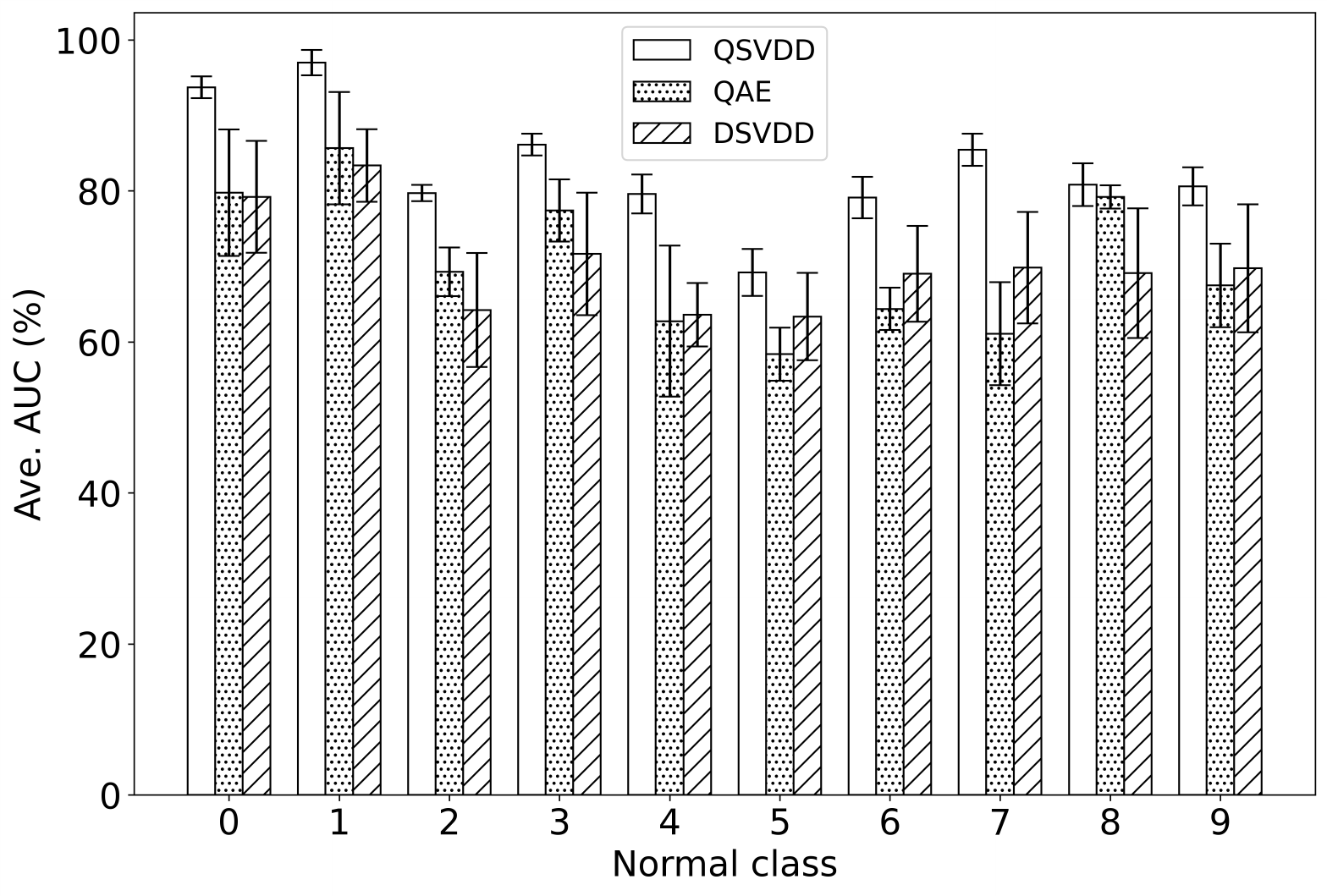}
  \captionsetup{margin = {0.4cm,0cm}}
  \caption{MNIST dataset}
  \label{Comparsion_MNIST-dataset}
\end{subfigure}
\begin{subfigure}[b]{.5\textwidth}
  \centering
  \includegraphics[height=5.5cm]{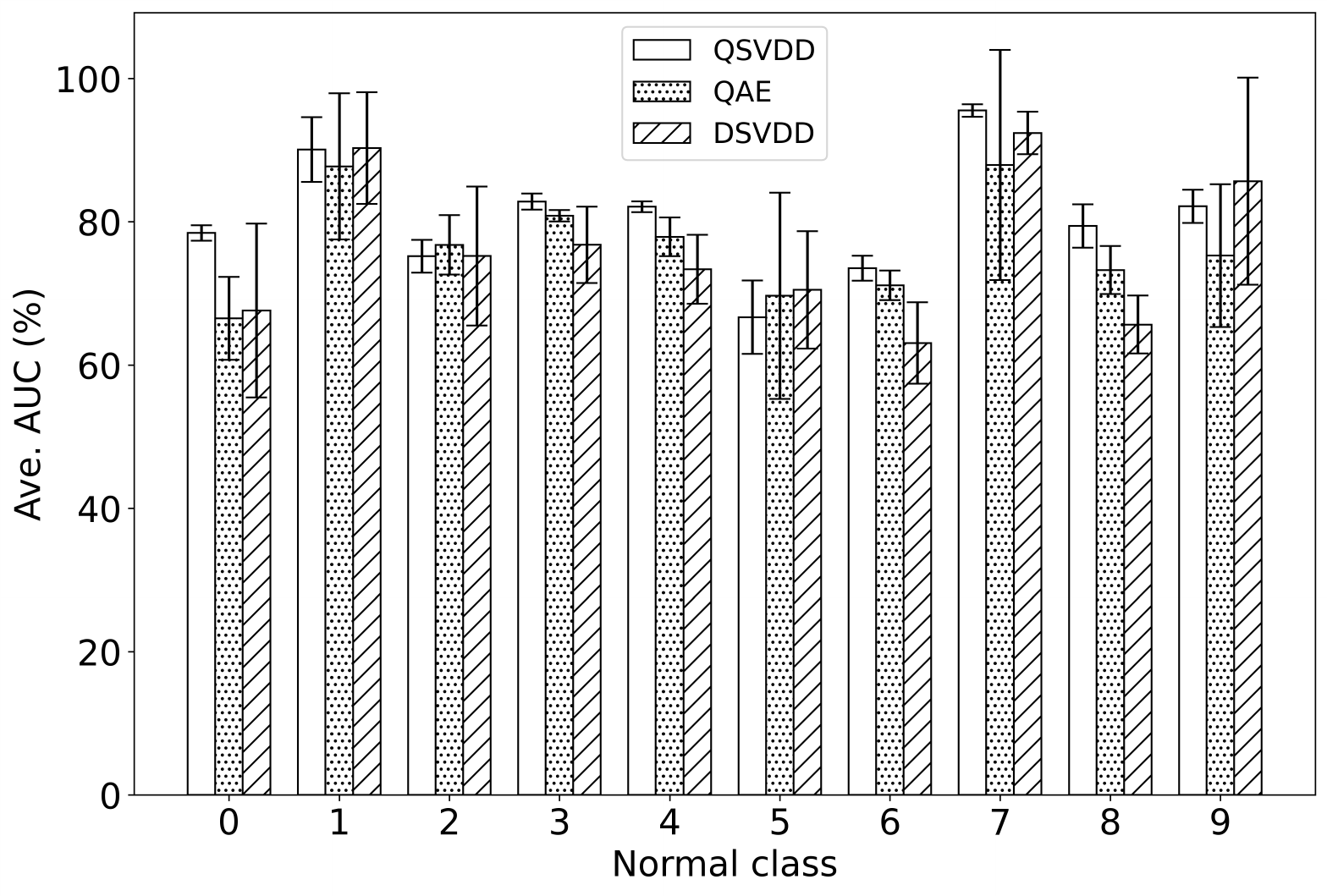}  
  \captionsetup{margin = {0.4cm,0cm}}
  \caption{Fashion MNIST dataset}
  \label{Comparison_Fashion-MNIST-dataset}
\end{subfigure}
\caption{The bar chart presents the average AUC scores and standard deviations achieved by QSVDD, QAE, and DSVDD under similar training conditions for (a) the MNIST dataset and (b) the Fashion MNIST dataset. In all cases, the latent space dimension is fixed at nine. For each dataset, ten independent AD tests were conducted, with each test designating one of the classes as the normal class. Each bar represents the average score computed from five repetitions of each AD test, with random initialization of parameters. Error bars indicate the standard deviations. The results demonstrate that QSVDD consistently outperforms the other methods for the MNIST dataset and in seven out of ten instances for the Fashion MNIST dataset. Additionally, the standard deviations in the QSVDD method consistently show smaller variations compared to the other methods, highlighting the stability of the proposed approach.}
\label{AUC_comparison}
\end{figure}

We compare QSVDD to the QAE-based OCC and DSVDD, each representing quantum and classical AD methods, respectively. The QAE-based OCC used in this comparative study is sourced from Refs.~\cite{bravo2021quantum,VQOCC}, and its details are provided in~\ref{appendix:QAE}. The structures of the VQC and neural network are set up in such a way that the number of parameters in the different methods is comparable. Specifically, QSVDD, QAE, and DSVDD are configured with 75, 78, and 92 parameters, respectively. The simulation results are presented in Fig.~\ref{AUC_comparison}. In this figure, the heights of the empty, dotted, and hatched bars represent the average AUCs in percentage obtained with QSVDD, QAE, and DSVDD, respectively, computed from five random seeds. The error bars indicate the standard deviations.

The simulation results demonstrate the strong performance of QSVDD. Specifically, QSVDD outperforms both QAE and DSVDD in all test cases where different normal classes are used for the MNIST dataset and in seven out of ten test cases for the Fashion-MNIST dataset.

For the MNIST dataset, QSVDD achieved a mean AUC of 83.13 when averaged over all normal classes, while QAE and DSVDD scored 70.54 and 70.32, respectively. Furthermore, QSVDD exhibited a smaller average standard deviation of 2.15 compared to 5.34 and 7.29 in QAE and DSVDD, respectively. The reduction in the standard deviation suggests that QSVDD is more reliable than the other two methods.

For the Fashion-MNIST dataset, QSVDD attained a mean AUC of 80.59 when averaged over all normal classes, while QAE and DSVDD scored 76.69 and 76.05, respectively. QSVDD once again demonstrated improved reliability, with an average standard deviation of 2.28, compared to 6.94 for QAE and 7.50 for DSVDD.

As described in Sec.~\ref{sec:qsvdd_vqc}, both the circuit depth and the number of parameters of the QSVDD with the QCNN ansatz grow logarithmically with the number of qubits. In contrast, the number of parameters grows linearly, and the circuit depth grows quadratically with the number of qubits in the QAE-based AD (see \ref{appendix:QAE}). Therefore, QSVDD emerges as a more efficient QML method for AD than QAE. Moreover, the reduction in the circuit depth is imperative from the practical perspective, as the size of the quantum circuit that can be executed reliably is limited in NISQ devices. It is also noteworthy that QSVDD outperformed DSVDD in most cases, despite using fewer model parameters.

\section{Conclusion}\label{sec:conc}
Anomaly detection aims to identify data points or patterns that deviate significantly from expected or normal behavior within a dataset, playing a crucial role in data science across various domains. This work presented the QSVDD algorithm, the quantum analogue of deep SVDD, for one-class classification and its application to anomaly detection. Our approach involves training a shallow-depth variational quantum circuit and projective measurements to learn a low-dimensional feature space in which the normal data is enclosed by the minimum-volume hypersphere. Utilizing QCNN as the trainable quantum circuit for QSVDD enables logarithmic growth in both the number of parameters and the quantum circuit depth as the number of qubits increases. This implies a doubly exponential reduction in the number of parameters and circuit depth relative to the number of data features within the input quantum state, if the exponentially large quantum state space is fully utilized. This compact design is also advantageous for implementation on near-term quantum devices, which rely on limited quantum resources.

Numerical simulations were conducted on 20 instances of AD tasks using the MNIST and Fashion MNIST image datasets. These simulations demonstrated that QSVDD outperformed both QAE, a quantum machine learning algorithm, and DSVDD, a deep learning algorithm. Notably, this improvement was achieved despite QSVDD having a shallower circuit depth than QAE and fewer parameters than DSVDD. These results strongly suggest that QSVDD holds great promise as a highly efficient quantum machine learning approach for AD. Its advantage over classical approaches is expected to become even more pronounced when applied to data that is inherently quantum, as produced by quantum devices such as quantum sensors~\cite{10.1038/s43588-022-00311-3}.

While the primary objective of this work is to establish a general framework for QSVDD based on VQC, the algorithm can be customized at various stages to accommodate specific tasks and data. For instance, when applying QSVDD to classical data, optimizing the data encoding method beyond amplitude encoding could result in a more efficient or practical quantum representation of the target input data. Further refinement in selecting the VQC structure, measurement scheme, and optimization algorithm could tailor the QSVDD algorithm to better suit other specific tasks and datasets. Furthermore, exploring the integration of general quantum measurements, such as the positive-operator valued measure (POVM), beyond projective measurements and optimizing it~\cite{lee2023variational}, represents an intriguing avenue for future research. Another intriguing extension of this work involves applying classical-to-quantum transfer learning~\cite{Mari2020transferlearningin,kim2023classical} to the domain of AD to enhance the utility of QML.



\section*{Acknowledgments}
This work was supported by Institute for Information \& communications Technology Promotion (IITP) grant funded by the Korea government(MSIP) (No. 2019-0-00003, Research and Development of Core technologies for Programming, Running, Implementing and Validating of Fault-Tolerant Quantum Computing System), the Yonsei University Research Fund of 2023 (2023-22-0072), the National Research Foundation of Korea (Grant No. 2022M3E4A1074591), and the KIST Institutional Program (2E32241-23-010). We thank Gunhee Park for helpful discussions.

\section*{Data availability}
The data that support the findings of this study are available upon request from the authors.

\appendix
\section{Quantum autoencoder}
\label{appendix:QAE}
\begin{figure*}[t]
    \centering
    \includegraphics[width=\linewidth]{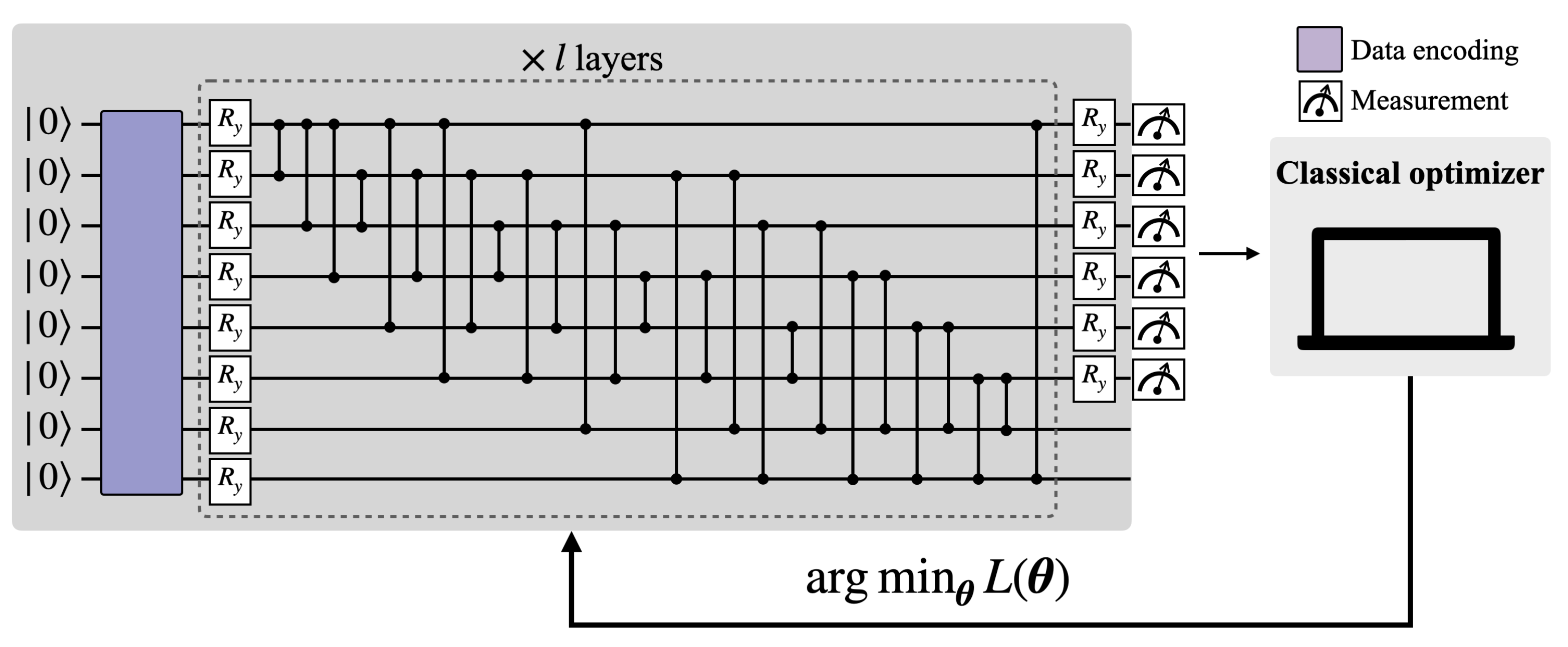}
    \caption{Illustration of the QAE ansatz used in this work for the benchmark study. In our implementation, we set $n=8$, $n_t=6$, and the last two qubits at the bottom represent the latent qubits of the QAE. The quantum state of the two latent qubits is projected onto the 9-dimensional latent (feature) space using the expectation measurements of 9 Pauli observables, similar to the QSVDD with the QCNN ansatz.
    }
    \label{fig: QAE}
\end{figure*}

Quantum autoencoder (QAE)~\cite{romero2017quantum} aims to learn a compressed representation of quantum data, ideally without losing information, by training a variational quantum circuit. In the QAE framework, the data originally represented by $n$ input qubits is compressed to $n-n_t$ latent qubits, where $n_t>0$ is the number of trash qubits that are eliminated. The training procedure for constructing a QAE is based on the following intuition: if the encoder transforms trash qubits into the state $|0\rangle^{\otimes n_t}$, then the decoder can perfectly reconstruct the original data by applying the inverse of the encoder to $n_t$ reference states initialized in $|0\rangle^{\otimes n_t}$ and the latent qubits. This also aligns with the goal that the trash qubits should not contain any useful information about the original state. The reduced density matrix describing the trash qubits after the parameterized encoding operator $U(\boldsymbol{\theta})$ can be expressed as $\rho_{\mathrm{trash}}(\boldsymbol{\theta})=\mathrm{Tr}_{\mathrm{latent}}\left(U(\boldsymbol{\theta})\rho(x)U^{\dagger}(\boldsymbol{\theta})\right)$, where $\rho(x)$ is the density matrix representation of the input data, and $\mathrm{Tr}_{\mathrm{latent}}(\cdot)$ denotes the partial trace performed on the latent qubits. The loss function is then expressed as
\begin{equation} \label{eq: qae}
L(\boldsymbol{\theta}) =  \sum_{j=1}^{n_t} (1- \mathrm{Tr}(\sigma_z^{(j)}\rho_{\mathrm{trash}}(\boldsymbol{\theta})),
\end{equation}
where $n_t$ is the number of trash qubits, and $\sigma_z^{(j)}$ is the Pauli $z$ observable whose expectation is measured on the $j^{\mathrm{th}}$ qubit. It is evident that this loss function is minimized when $\rho_{\mathrm{trash}}(\boldsymbol{\theta})=(|0\rangle\langle 0 |)^{\otimes n_t}$. Inspired by the loss function shown in Eq.~(\ref{eq: qae}), Refs.~\cite{bravo2021quantum, VQOCC} proposed an ansatz designed to disentangle the trash qubits from each other, as well as from the latent qubits. The encoding circuit inspired by this principle, with $n=8$ and $n_t=6$, is illustrated in Fig.~\ref{fig: QAE}. The trainable parameters are the angles of the single-qubit rotation around the $y$-axis of the Bloch sphere, denoted as $R_y$ in the figure.

Several studies have demonstrated that QAE can effectively address AD problems~\cite{AD_qae, AD_qae_hybrid, VQOCC}. When the primary goal is anomaly detection, there is no need to reconstruct the original data through decoding. Consequently, the quantum circuit depth is reduced by half. We adopted the QAE-based one-class classification approach proposed in Ref.~\cite{VQOCC} and compared its performance with our QSVDD algorithm. In our study, the QAE with $n=8$ and $n_t=6$ as depicted in Fig.~\ref{fig: QAE} is trained to minimize the loss function shown in Eq.~(\ref{eq: qae}). Subsequently, the quantum state of the two latent qubits is projected onto the 9-dimensional subspace by the Pauli measurements, as outlined in Section~\ref{sec:qsvdd_meas}. The center $\boldsymbol{c}$ and the radius $r$ of the hypersphere can then be determined as the mean of the normal data and the largest distance between the normal data from the center, respectively. Finally, the decision function shown in Eq.~(\ref{eq: decision_QSVDD}) is used to identify anomalies.

To compare the AD performance of the QAE-based algorithm and our QSVDD under similar training conditions, we designed the QAE ansatz to employ 78 trainable parameters, while the QSVDD with the QCNN ansatz employes 75 parameters. In general, the number of parameters and the circuit depth of the QSVDD with the QCNN ansatz grow logarithmically with the number of input qubits. In constrast, in QAE, the number of parameters ($p$) scales as
\begin{equation} \label{eq: qae_params}
p = n_t + nl,
\end{equation}
and the circuit depth ($d_c$) scales as
\begin{equation} \label{eq: qae_depth}
d_c = 1 + \Bigg( {n_t \choose 2} + n_t(n-n_t) + 1 \Bigg)l,
\end{equation}
where $l$ is the number of layers used in QAE. We can assume that $n_t = n/c$, where $c>0$ is a constant, meaning that a fraction of the input qubits is used to store the compressed information. Therefore, $p\in O(n)$ and $d_c\in O(n^2)$.

\section{QSVDD simulation results}
\label{appendix: QSVDD_results}

In Section~\ref{sec:numexp}, we presented simulation results illustrating the performance of QSVDD. This appendix contains the complete numerical values of these QSVDD simulation results.
\begin{table*}[!htb]
\centering

\begin{tabular}{ll|c|c|c|c|c|c}
\hline\hline

\multicolumn{2}{c|}{\multirow{2}{*}{Normal class}} & \multicolumn{6}{c}{Dimension of latent space} \cr
\cline{3-8}
& & 1 & 3 & 6 & 9 & 12 & 15\\

\hline
& 0 & 
79.55 $\pm$ 4.11&
90.73 $\pm$ 2.65&
91.08 $\pm$ 1.67&
93.73 $\pm$ 1.44&
94.47 $\pm$ 0.98&
94.29 $\pm$ 1.26    \\

& 1 & 
85.89 $\pm$ 2.34&
95.85 $\pm$ 1.12&
97.31 $\pm$ 0.73&
96.99 $\pm$ 1.68&
97.97 $\pm$ 0.43&
97.94 $\pm$ 0.18    \\

& 2 & 
68.12 $\pm$ 3.86&
74.72 $\pm$ 1.46&
74.88 $\pm$ 7.71&
79.71 $\pm$ 1.08&
76.78 $\pm$ 4.98&
78.65 $\pm$ 2.56    \\
                               
& 3 & 
70.46 $\pm$ 3.98&
80.34 $\pm$ 1.64&
84.37 $\pm$ 2.38&
86.13 $\pm$ 1.44&
84.98 $\pm$ 3.21&
84.89 $\pm$ 2.52   \\
                               
& 4 & 
68.36 $\pm$ 5.48&
79.18 $\pm$ 3.88&
82.84 $\pm$ 4.04&
79.59 $\pm$ 2.58&
79.45 $\pm$ 5.08&
81.24 $\pm$ 3.57   \\
                               
& 5 & 
62.90 $\pm$ 2.13&
66.07 $\pm$ 3.46&
70.2 $\pm$ 3.46&
69.2 $\pm$ 3.11&
68.68 $\pm$ 4.73&
68.19 $\pm$ 2.65    \\
                               
& 6 & 
67.57 $\pm$ 5.26&
74.92 $\pm$ 3.78&
76.86 $\pm$ 2.97&
79.11 $\pm$ 2.74&
79.09 $\pm$ 1.77&
77.37 $\pm$ 0.81   \\
                               
& 7 & 
75.36 $\pm$ 4.66&
86.28 $\pm$ 1.95&
86.19 $\pm$ 2.59&
85.45 $\pm$ 2.14&
85.11 $\pm$ 1.47&
86.52 $\pm$ 1.9    \\
                               
& 8 & 
68.88 $\pm$ 2.83&
76.33 $\pm$ 2.37&
80.06 $\pm$ 3.86&
80.83 $\pm$ 2.83&
82.08 $\pm$ 2.0&
82.17 $\pm$ 2.87    \\
                               
& 9 & 
69.74 $\pm$ 3.96&
74.24 $\pm$ 2.03&
78.34 $\pm$ 3.65&
80.6 $\pm$ 2.51&
77.41 $\pm$ 4.39&
75.84 $\pm$ 1.65   \\

& Mean & 
71.68 $\pm$ 3.86&
79.87 $\pm$ 2.43&
82.21 $\pm$ 3.31&
83.13 $\pm$ 2.16&
82.60 $\pm$ 2.90&
82.71 $\pm$ 2.00   \\
            
\hline

& T-shirt/top &
67.36 $\pm$ 4.16&
74.52 $\pm$ 1.53&
76.72 $\pm$ 1.45&
78.43 $\pm$ 1.08&
78.17 $\pm$ 2.41&
77.98 $\pm$ 0.90	\\
& Trouser &
80.65 $\pm$ 4.27&
87.37 $\pm$ 2.84&
91.44 $\pm$ 3.18&
90.09 $\pm$ 4.52&
90.78 $\pm$ 3.32&
87.64 $\pm$ 2.86	\\
& Pullover &
67.91 $\pm$ 1.46&
71.71 $\pm$ 2.07&
73.21 $\pm$ 2.28&
75.18 $\pm$ 2.29&
73.70 $\pm$ 1.72&
74.37 $\pm$ 1.13	\\
& Dress &
71.49 $\pm$ 1.46&
79.08 $\pm$ 1.59&
81.64 $\pm$ 1.42&
82.81 $\pm$ 1.13&
82.04 $\pm$ 0.78&
82.69 $\pm$ 0.92	\\
& Coat &
74.43 $\pm$ 1.96&
79.15 $\pm$ 0.57&
81.22 $\pm$ 2.04&
82.09 $\pm$ 0.75&
81.42 $\pm$ 1.82&
82.42 $\pm$ 1.03	\\
& Sandal &
55.77 $\pm$ 6.38&
69.63 $\pm$ 6.01&
62.07 $\pm$ 5.91&
66.67 $\pm$ 5.11&
67.78 $\pm$ 7.98&
61.42 $\pm$ 5.14	\\
& Shirt &
65.82 $\pm$ 1.91&
73.32 $\pm$ 1.74&
74.45 $\pm$ 1.34&
73.50 $\pm$ 1.74&
75.43 $\pm$ 1.32&
74.17 $\pm$ 0.69	\\
& Sneaker &
94.78 $\pm$ 0.75&
93.79 $\pm$ 1.31&
94.53 $\pm$ 1.85&
95.54 $\pm$ 0.87&
96.17 $\pm$ 1.19&
97.02 $\pm$ 0.67	\\
& Bag &
72.04 $\pm$ 6.36&
68.80 $\pm$ 6.16&
76.79 $\pm$ 5.48&
79.41 $\pm$ 3.03&
74.35 $\pm$ 4.32&
73.11 $\pm$ 4.63	\\
& Ankle boot &
78.59 $\pm$ 3.46&
78.83 $\pm$ 3.84&
82.47 $\pm$ 3.45&
82.15 $\pm$ 2.32&
85.29 $\pm$ 1.71&
77.97 $\pm$ 1.46	\\

& Mean &
72.88 $\pm$ 3.22&
77.62 $\pm$ 2.77&
79.45 $\pm$ 2.84&
80.59 $\pm$ 2.28&
80.51 $\pm$ 2.66&
78.88 $\pm$ 1.94	\\
            
\hline\hline

\end{tabular}
\caption{The table presents the average AUC (\%) and standard deviations obtained by the QSVDD algorithm in five repeated experiments, each with random parameter initialization. The experiments were conducted on both the MNIST and Fashion MNIST datasets. Each column in the table corresponds to the dimension of the latent space, ranging from 1 to 15. Each row in the table represents the normal class in the anomaly detection problem. For a visual representation of this table, see Figure~\ref{AUC_mean}.
}
\label{total_cell}
\end{table*}


\begin{thebibliography}{10}

\bibitem{10.1038/s43588-022-00311-3}
M.~Cerezo, Guillaume Verdon, Hsin-Yuan Huang, Lukasz Cincio, and Patrick~J.
  Coles.
\newblock {Challenges and opportunities in quantum machine learning}.
\newblock {\em Nature Computational Science}, pages 1--10, 2022.

\bibitem{10.1080/00107514.2014.964942}
Maria Schuld, Ilya Sinayskiy, and Francesco Petruccione.
\newblock {An introduction to quantum machine learning}.
\newblock {\em Contemporary Physics}, 56(2):172--185, 2015.

\bibitem{10.1038/nature23474}
Jacob Biamonte, Peter Wittek, Nicola Pancotti, Patrick Rebentrost, Nathan
  Wiebe, and Seth Lloyd.
\newblock {Quantum machine learning}.
\newblock {\em Nature}, 549(7671):195--202, 2017.

\bibitem{10.1103/physrevlett.113.130503}
Patrick Rebentrost, Masoud Mohseni, and Seth Lloyd.
\newblock {Quantum Support Vector Machine for Big Data Classification}.
\newblock {\em Physical Review Letters}, 113(13):130503, 2014.

\bibitem{10.1088/2058-9565/ab5944}
Marcello Benedetti, Erika Lloyd, Stefan Sack, and Mattia Fiorentini.
\newblock {Parameterized quantum circuits as machine learning models}.
\newblock {\em Quantum Science and Technology}, 5(1):019601, 2020.

\bibitem{cerezo2020variational}
M.~Cerezo, Andrew Arrasmith, Ryan Babbush, Simon~C. Benjamin, Suguru Endo,
  Keisuke Fujii, Jarrod~R. {McClean}, Kosuke Mitarai, Xiao Yuan, Lukasz Cincio,
  and Patrick~J. Coles.
\newblock Variational quantum algorithms.
\newblock {\em Nature Reviews Physics}, 3(9):625--644, 2021.

\bibitem{abbas_power_2021}
Amira Abbas, David Sutter, Christa Zoufal, Aurelien Lucchi, Alessio Figalli,
  and Stefan Woerner.
\newblock The power of quantum neural networks.
\newblock {\em Nature Computational Science}, 1(6):403--409, 2021.

\bibitem{phua2010comprehensive}
Clifton Phua, Vincent Lee, Kate Smith, and Ross Gayler.
\newblock A comprehensive survey of data mining-based fraud detection research.
\newblock {\em arXiv preprint arXiv:1009.6119}, 2010.

\bibitem{li2012identifying}
Shing-Han Li, David~C Yen, Wen-Hui Lu, and Chiang Wang.
\newblock Identifying the signs of fraudulent accounts using data mining
  techniques.
\newblock {\em Computers in Human Behavior}, 28(3):1002--1013, 2012.

\bibitem{jeragh2018combining}
Mohamad Jeragh and Mousa AlSulaimi.
\newblock Combining auto encoders and one class support vectors machine for
  fraudulant credit card transactions detection.
\newblock In {\em 2018 Second World Conference on Smart Trends in Systems,
  Security and Sustainability (WorldS4)}, pages 178--184. IEEE, 2018.

\bibitem{feher2014cell}
Kristen Feher, Jenny Kirsch, Andreas Radbruch, Hyun-Dong Chang, and Toralf
  Kaiser.
\newblock Cell population identification using fluorescence-minus-one controls
  with a one-class classifying algorithm.
\newblock {\em Bioinformatics}, 30(23):3372--3378, 2014.

\bibitem{min2017deep}
Seonwoo Min, Byunghan Lee, and Sungroh Yoon.
\newblock Deep learning in bioinformatics.
\newblock {\em Briefings in bioinformatics}, 18(5):851--869, 2017.

\bibitem{marti2015anomaly}
Luis Mart{\'\i}, Nayat Sanchez-Pi, Jos{\'e}~Manuel Molina, and Ana
  Cristina~Bicharra Garcia.
\newblock Anomaly detection based on sensor data in petroleum industry
  applications.
\newblock {\em Sensors}, 15(2):2774--2797, 2015.

\bibitem{6618951}
Babak Saleh, Ali Farhadi, and Ahmed Elgammal.
\newblock Object-centric anomaly detection by attribute-based reasoning.
\newblock In {\em 2013 IEEE Conference on Computer Vision and Pattern
  Recognition}, pages 787--794, 2013.

\bibitem{bao2019computer}
Yuequan Bao, Zhiyi Tang, Hui Li, and Yufeng Zhang.
\newblock Computer vision and deep learning--based data anomaly detection
  method for structural health monitoring.
\newblock {\em Structural Health Monitoring}, 18(2):401--421, 2019.

\bibitem{Fraser2022}
Katherine Fraser, Samuel Homiller, Rashmish~K. Mishra, Bryan Ostdiek, and
  Matthew~D. Schwartz.
\newblock Challenges for unsupervised anomaly detection in particle physics.
\newblock {\em Journal of High Energy Physics}, 2022(3):66, Mar 2022.

\bibitem{chandola2009anomaly}
Varun Chandola, Arindam Banerjee, and Vipin Kumar.
\newblock Anomaly detection: A survey.
\newblock {\em ACM computing surveys (CSUR)}, 41(3):1--58, 2009.

\bibitem{chalapathy2019deep}
Raghavendra Chalapathy and Sanjay Chawla.
\newblock Deep learning for anomaly detection: A survey.
\newblock {\em arXiv preprint arXiv:1901.03407}, 2019.

\bibitem{MOYA1996463}
Mary~M. Moya and Don~R. Hush.
\newblock Network constraints and multi-objective optimization for one-class
  classification.
\newblock {\em Neural Networks}, 9(3):463--474, 1996.

\bibitem{OCC_survey}
Shehroz~S Khan and Michael~G Madden.
\newblock One-class classification: taxonomy of study and review of techniques.
\newblock {\em The Knowledge Engineering Review}, 29(3):345--374, 2014.

\bibitem{OCC_survey2}
Pramuditha Perera, Poojan Oza, and Vishal~M Patel.
\newblock {One-Class Classification: A Survey}.
\newblock {\em arXiv}, 2021.

\bibitem{OC-SVM}
Bernhard Sch{\"o}lkopf, Robert~C Williamson, Alex Smola, John Shawe-Taylor, and
  John Platt.
\newblock Support vector method for novelty detection.
\newblock {\em Advances in neural information processing systems}, 12, 1999.

\bibitem{SVDD}
David~MJ Tax and Robert~PW Duin.
\newblock Support vector data description.
\newblock {\em Machine learning}, 54:45--66, 2004.

\bibitem{DSVDD}
Lukas Ruff, Robert Vandermeulen, Nico Goernitz, Lucas Deecke, Shoaib~Ahmed
  Siddiqui, Alexander Binder, Emmanuel M{\"u}ller, and Marius Kloft.
\newblock Deep one-class classification.
\newblock In {\em International conference on machine learning}, pages
  4393--4402. PMLR, 2018.

\bibitem{OCC_survey_deep}
Raghavendra Chalapathy and Sanjay Chawla.
\newblock Deep learning for anomaly detection: A survey.
\newblock {\em arXiv preprint arXiv:1901.03407}, 2019.

\bibitem{QML_for_AD}
Nana Liu and Patrick Rebentrost.
\newblock Quantum machine learning for quantum anomaly detection.
\newblock {\em Phys. Rev. A}, 97:042315, Apr 2018.

\bibitem{QAD_audio}
Zihua Chai, Ying Liu, Mengqi Wang, Yuhang Guo, Fazhan Shi, Zhaokai Li, Ya~Wang,
  and Jiangfeng Du.
\newblock Quantum anomaly detection of audio samples with a spin processor in
  diamond.
\newblock {\em arXiv preprint arXiv:2201.10263}, 2022.

\bibitem{QAD_phase_diagram}
Korbinian Kottmann, Friederike Metz, Joana Fraxanet, and Niccol\`o Baldelli.
\newblock Variational quantum anomaly detection: Unsupervised mapping of phase
  diagrams on a physical quantum computer.
\newblock {\em Phys. Rev. Research}, 3:043184, Dec 2021.

\bibitem{VQOCC}
Gunhee Park, Joonsuk Huh, and Daniel~Kyungdeock Park.
\newblock Variational quantum one-class classifier.
\newblock {\em Machine Learning: Science and Technology}, 2022.

\bibitem{HHL}
Aram~W. Harrow, Avinatan Hassidim, and Seth Lloyd.
\newblock Quantum algorithm for linear systems of equations.
\newblock {\em Phys. Rev. Lett.}, 103:150502, Oct 2009.

\bibitem{qPCA}
Seth Lloyd, Masoud Mohseni, and Patrick Rebentrost.
\newblock Quantum principal component analysis.
\newblock {\em Nature Physics}, 10(9):631--633, 2014.

\bibitem{NISQ}
John Preskill.
\newblock Quantum {C}omputing in the {NISQ} era and beyond.
\newblock {\em {Quantum}}, 2:79, August 2018.

\bibitem{romero2017quantum}
Jonathan Romero, Jonathan~P Olson, and Alan Aspuru-Guzik.
\newblock Quantum autoencoders for efficient compression of quantum data.
\newblock {\em Quantum Science and Technology}, 2(4):045001, 2017.

\bibitem{bravo2021quantum}
Carlos Bravo-Prieto.
\newblock Quantum autoencoders with enhanced data encoding.
\newblock {\em Machine Learning: Science and Technology}, 2(3):035028, 2021.

\bibitem{AD_qae}
Vishal~S Ngairangbam, Michael Spannowsky, and Michihisa Takeuchi.
\newblock Anomaly detection in high-energy physics using a quantum autoencoder.
\newblock {\em Physical Review D}, 105(9):095004, 2022.

\bibitem{AD_qae_hybrid}
Alona Sakhnenko, Corey O’Meara, Kumar~JB Ghosh, Christian~B Mendl, Giorgio
  Cortiana, and Juan Bernab{\'e}-Moreno.
\newblock Hybrid classical-quantum autoencoder for anomaly detection.
\newblock {\em Quantum Machine Intelligence}, 4(2):27, 2022.

\bibitem{cerezo2021variational}
Marco Cerezo, Andrew Arrasmith, Ryan Babbush, Simon~C Benjamin, Suguru Endo,
  Keisuke Fujii, Jarrod~R McClean, Kosuke Mitarai, Xiao Yuan, Lukasz Cincio,
  et~al.
\newblock Variational quantum algorithms.
\newblock {\em Nature Reviews Physics}, 3(9):625--644, 2021.

\bibitem{blance2021quantum}
Andrew Blance and Michael Spannowsky.
\newblock Quantum machine learning for particle physics using a variational
  quantum classifier.
\newblock {\em Journal of High Energy Physics}, 2021(2):1--20, 2021.

\bibitem{chen2020variational}
Samuel Yen-Chi Chen, Chao-Han~Huck Yang, Jun Qi, Pin-Yu Chen, Xiaoli Ma, and
  Hsi-Sheng Goan.
\newblock Variational quantum circuits for deep reinforcement learning.
\newblock {\em IEEE Access}, 8:141007--141024, 2020.

\bibitem{romero2021variational}
Jonathan Romero and Al{\'a}n Aspuru-Guzik.
\newblock Variational quantum generators: Generative adversarial quantum
  machine learning for continuous distributions.
\newblock {\em Advanced Quantum Technologies}, 4(1):2000003, 2021.

\bibitem{cong_QCNN}
Iris Cong, Soonwon Choi, and Mikhail~D. Lukin.
\newblock Quantum convolutional neural networks.
\newblock {\em Nature Physics}, 15(12):1273--1278, December 2019.

\bibitem{pesah2020absence}
Arthur Pesah, M.~Cerezo, Samson Wang, Tyler Volkoff, Andrew~T. Sornborger, and
  Patrick~J. Coles.
\newblock Absence of barren plateaus in quantum convolutional neural networks.
\newblock {\em Phys. Rev. X}, 11:041011, Oct 2021.

\bibitem{PRXQuantum.2.040321}
Leonardo Banchi, Jason Pereira, and Stefano Pirandola.
\newblock Generalization in quantum machine learning: A quantum information
  standpoint.
\newblock {\em PRX Quantum}, 2:040321, Nov 2021.

\bibitem{hur_QCNN}
Tak Hur, Leeseok Kim, and Daniel~K Park.
\newblock {Quantum convolutional neural network for classical data
  classification}.
\newblock {\em Quantum Machine Intelligence}, 4(1):3, 2022.

\bibitem{kim2023classical}
Juhyeon Kim, Joonsuk Huh, and Daniel~K. Park.
\newblock Classical-to-quantum convolutional neural network transfer learning.
\newblock {\em Neurocomputing}, 555:126643, 2023.

\bibitem{bergholm2020pennylane}
Ville Bergholm, Josh Izaac, Maria Schuld, Christian Gogolin, M.~Sohaib Alam,
  Shahnawaz Ahmed, Juan~Miguel Arrazola, Carsten Blank, Alain Delgado, Soran
  Jahangiri, Keri McKiernan, Johannes~Jakob Meyer, Zeyue Niu, Antal Száva, and
  Nathan Killoran.
\newblock Pennylane: Automatic differentiation of hybrid quantum-classical
  computations, 2020.

\bibitem{jain2000statistical}
Anil~K Jain, Robert P.~W. Duin, and Jianchang Mao.
\newblock Statistical pattern recognition: A review.
\newblock {\em IEEE Transactions on pattern analysis and machine intelligence},
  22(1):4--37, 2000.

\bibitem{SVM}
Corinna Cortes and Vladimir Vapnik.
\newblock Support-vector networks.
\newblock {\em Machine learning}, 20:273--297, 1995.

\bibitem{hofmann2006support}
Martin Hofmann.
\newblock Support vector machines-kernels and the kernel trick.
\newblock {\em Notes}, 26(3):1--16, 2006.

\bibitem{tax2002one}
David Martinus~Johannes Tax.
\newblock One-class classification: Concept learning in the absence of
  counter-examples.
\newblock 2002.

\bibitem{10.1162/089976601750264965}
Bernhard Schölkopf, John~C. Platt, John Shawe-Taylor, Alex~J. Smola, and
  Robert~C. Williamson.
\newblock {Estimating the Support of a High-Dimensional Distribution}.
\newblock {\em Neural Computation}, 13(7):1443--1471, 07 2001.

\bibitem{smola1998learning}
Alex~J Smola and Bernhard Sch{\"o}lkopf.
\newblock {\em Learning with kernels}, volume~4.
\newblock Citeseer, 1998.

\bibitem{pal2010feature}
Mahesh Pal and Giles~M Foody.
\newblock Feature selection for classification of hyperspectral data by svm.
\newblock {\em IEEE Transactions on Geoscience and Remote Sensing},
  48(5):2297--2307, 2010.

\bibitem{vempati2010generalized}
Sreekanth Vempati, Andrea Vedaldi, Andrew Zisserman, and CV~Jawahar.
\newblock Generalized rbf feature maps for efficient detection.
\newblock In {\em BMVC}, pages 1--11, 2010.

\bibitem{havlivcek2019supervised}
Vojt{\v{e}}ch Havl{\'\i}{\v{c}}ek, Antonio~D C{\'o}rcoles, Kristan Temme,
  Aram~W Harrow, Abhinav Kandala, Jerry~M Chow, and Jay~M Gambetta.
\newblock Supervised learning with quantum-enhanced feature spaces.
\newblock {\em Nature}, 567(7747):209--212, 2019.

\bibitem{schuld2019quantum}
Maria Schuld and Nathan Killoran.
\newblock Quantum machine learning in feature hilbert spaces.
\newblock {\em Physical review letters}, 122(4):040504, 2019.

\bibitem{Huang_2021}
Hsin-Yuan Huang, Michael Broughton, Masoud Mohseni, Ryan Babbush, Sergio Boixo,
  Hartmut Neven, and Jarrod~R. McClean.
\newblock Power of data in quantum machine learning.
\newblock {\em Nature Communications}, 12(1), may 2021.

\bibitem{PhysRevA.101.032308}
Maria Schuld, Alex Bocharov, Krysta~M. Svore, and Nathan Wiebe.
\newblock Circuit-centric quantum classifiers.
\newblock {\em Phys. Rev. A}, 101:032308, Mar 2020.

\bibitem{PhysRevA.102.032420}
Ryan LaRose and Brian Coyle.
\newblock Robust data encodings for quantum classifiers.
\newblock {\em Phys. Rev. A}, 102:032420, Sep 2020.

\bibitem{araujo_divide-and-conquer_2021}
Israel~F. Araujo, Daniel~K. Park, Francesco Petruccione, and Adenilton~J.
  da~Silva.
\newblock A divide-and-conquer algorithm for quantum state preparation.
\newblock {\em Scientific Reports}, 11(1):6329, 2021.

\bibitem{araujo2021configurable}
Israel~F. Araujo, Daniel~K. Park, Teresa~B. Ludermir, Wilson~R. Oliveira,
  Francesco Petruccione, and Adenilton~J. da~Silva.
\newblock Configurable sublinear circuits for quantum state preparation.
\newblock {\em Quantum Information Processing}, 22(2):123, 2023.

\bibitem{lloyd2020quantum}
Seth Lloyd, Maria Schuld, Aroosa Ijaz, Josh Izaac, and Nathan Killoran.
\newblock Quantum embeddings for machine learning.
\newblock {\em arXiv preprint arXiv:2001.03622}, 2020.

\bibitem{McClean2018}
Jarrod~R. McClean, Sergio Boixo, Vadim~N. Smelyanskiy, Ryan Babbush, and
  Hartmut Neven.
\newblock Barren plateaus in quantum neural network training landscapes.
\newblock {\em Nature Communications}, 9(1):4812, 2018.

\bibitem{maccormack_branching_2020}
Ian MacCormack, Conor Delaney, Alexey Galda, Nidhi Aggarwal, and Prineha
  Narang.
\newblock Branching quantum convolutional neural networks.
\newblock {\em Phys. Rev. Research}, 4:013117, Feb 2022.

\bibitem{liu2017survey}
Weibo Liu, Zidong Wang, Xiaohui Liu, Nianyin Zeng, Yurong Liu, and Fuad~E
  Alsaadi.
\newblock A survey of deep neural network architectures and their applications.
\newblock {\em Neurocomputing}, 234:11--26, 2017.

\bibitem{grant_hierarchical_2018}
Edward Grant, Marcello Benedetti, Shuxiang Cao, Andrew Hallam, Joshua Lockhart,
  Vid Stojevic, Andrew~G. Green, and Simone Severini.
\newblock Hierarchical quantum classifiers.
\newblock {\em npj Quantum Information}, 4(1):65, December 2018.

\bibitem{han2013comparison}
Dianyuan Han.
\newblock Comparison of commonly used image interpolation methods.
\newblock In {\em Conference of the 2nd International Conference on Computer
  Science and Electronics Engineering (ICCSEE 2013)}, pages 1556--1559.
  Atlantis Press, 2013.

\bibitem{kingma2014adam}
Diederik~P Kingma and Jimmy Ba.
\newblock Adam: A method for stochastic optimization.
\newblock {\em arXiv preprint arXiv:1412.6980}, 2014.

\bibitem{lee2023variational}
Dongkeun Lee, Kyunghyun Baek, Joonsuk Huh, and Daniel~K Park.
\newblock Variational quantum state discriminator for supervised machine
  learning.
\newblock {\em arXiv preprint arXiv:2303.03588}, 2023.

\bibitem{Mari2020transferlearningin}
Andrea Mari, Thomas~R. Bromley, Josh Izaac, Maria Schuld, and Nathan Killoran.
\newblock Transfer learning in hybrid classical-quantum neural networks.
\newblock {\em {Quantum}}, 4:340, October 2020.

\end{thebibliography}

\end{document}